\documentclass[12pt]{article}
\usepackage{graphics}
\usepackage{epsfig}
\begin{document}
\newcommand{\beq}{\begin{equation}}
\newcommand{\eeq}{\end{equation}}
\newcommand{\bea}{\begin{eqnarray}}
\newcommand{\eea}{\end{eqnarray}}
\newcommand{\eps}{\varepsilon}
\newcommand{\Fs}{\mbox{\scriptsize F}}
\newcommand{\lsim}{\stackrel{\scriptstyle <}{\phantom{}_{\sim}}}
\newcommand{\gsim}{\stackrel{\scriptstyle >}{\phantom{}_{\sim}}}

{\bf The role of the boundary conditions in the Wigner-Seitz
approximation applied to the neutron star inner crust.}
 \vskip 0.5 cm
\centerline{ M.~Baldo$^{a}$, E.E.~Saperstein$^{b}$ and
S.V.~Tolokonnikov$^{b}$ }
 \vskip 0.5 cm \centerline{$^a$INFN,
Sezione di Catania, 64 Via S.-Sofia, I-95123 Catania, Italy}
\centerline {$^{b}$ Kurchatov Institute, 123182, Moscow, Russia }
 \vskip 0.5 cm
\begin{abstract}
The influence of the boundary conditions used in the Wigner-Seitz
(WS) approximation applied to the neutron star inner crust is
examined. The generalized energy functional method which includes
the neutron and proton pairing correlations is used. Predictions of
two versions of the boundary conditions are compared with each
other. The uncertainties in the equilibrium configuration ($Z,R_c$)
of the crust, where $Z$ is the proton charge and $R_c$, the radius
of the WS cell, correspond to variation of $Z$ by 2 -- 6 units and
of $R_c$, by 1 -- 2 fm. The effect of the boundary conditions is
enhanced at increasing density. These uncertainties are smaller than
the variation of $Z$ and $R_c$ coming from the pairing effects.
 In the case of high densities, $k_{\Fs} \gsim 1\;$fm$^{-1}$, the
 most important uncertainty occurs in the value of the neutron gap
 $\Delta_n$. In the WS approximation, it originates from the shell
effect in the neutron single-particle spectrum which is rather
pronounced in the case of larger $k_{\Fs}$ and, correspondingly,
small $R_c$ values, but it becomes negligible at lower density near
the drip point. An approximate recipe to avoid this uncertainty is
suggested.
\end{abstract}

\noindent
PACS : 26.60.+c,97.60.Jd,21.65.+f,21.60.-n,21.30.Fe

\vskip 0.4 cm
\section{Introduction}
In the last two decades the interest on the structure of the neutron
star inner crust has been stimulated by the increasing number of
observational data on the pulsar glitches. The latter are commonly
explained in terms of the dynamics of superfluid vortices within the
inner crust of neutron stars (see \cite{Pet} and Refs. therein). By
``inner crust'' one usually indicates the part of the shell of a
neutron star with sub-nuclear densities $0.001 \rho_0 \lsim \rho
\lsim 0.5 \rho_0 $, where $\rho_0$ is the normal nuclear density.
 According to present-day ideas, the bulk
of the inner crust consists mainly of spherically symmetrical
nuclear-like clusters which form a crystal matrix immersed in a sea
of neutrons and virtually uniform sea of electrons.
 Such a picture was first justified microscopically in the classical
paper by Negele and Vautherin  \cite{NV} within the Wigner-Seitz
(WS) approximation. This approximation consists in replacing the
crystal by a sum  of identical spherical cells with a nuclear
cluster at the cell center. As far as the WS  method is inconsistent
with the crystal symmetry  (periodicity), it
 implies a need for some artificial boundary conditions  for the
 neutron wave functions at the  WS  cell boundary. To some extent,
these conditions are arbitrary provided they guarantee the
orthogonality and completeness of the single-particle basis.
 Unfortunately, calculations with complete
taking into account the crystal symmetry are quite complicated.
There are only  few such calculations, being limited mainly with
consideration of the deep (high density) layers of the
crust\cite{Mag,CCH}, where the ``lasagna'' or ``spaguetti''
structure of the crust matter is assumed and the use of the WS
method with the spherical symmetry cannot be applied. Only recently
a consistent band theory was developed for the outer (low density)
layers as well \cite{NC}. Therefore up to now  the WS method is
quite popular and it is usually considered as the most practical one
for systematic investigation of the inner crust structure in the
whole density interval. For describing the matter of a neutron star
crust, Negele and Vautherin used a version of the energy functional
method with the density dependent effective mass $m^*(\rho)$. In
fact, it is very close to the Hartree-Fock method with effective
Skyrme forces. For a fixed average nuclear density $\rho$, the
nuclear (plus electron) energy functional is minimized for the
spherical WS cell of the radius $R_c$. A cell contains $Z$ protons
(and electrons) and $N{=}A{-}Z$ neutrons ($A=(4\pi/3)R_c^3\rho$). In
addition, the $\beta$-stability condition,
 \beq
 \mu_n-(\mu_p+\mu_e)=0, \label{mu}
 \eeq
has to be fulfilled, where $\mu_n$, $\mu_p$ and $\mu_e$ are the
chemical potentials of neutrons, protons and electrons,
respectively. The minimization procedure is carried out for
different values of $Z$ and $R_c$. The equilibrium configuration
($Z,R_c$) at the considered density corresponds to the absolute
minimum in energy among all these possible configurations.
Application of the variational principle to the  energy functional
by Negele and Vautherin for a WS cell results in a set of the
Shr\"odinger-type equations for the single particle neutron
functions $\phi_{\lambda}({\bf r})=R_{nlj}(r) \Phi_{ljm}({\bf
n})$, with the standard notation. The radial functions
$R_{nlj}(r)$ obey the boundary condition  at the point $r=R_c$. As
it was noted above, different kinds of boundary conditions could
be used. Negele and Vautherin used the following one:
 \beq R_{nlj}(r=R_c)=0
 \label{bco}
\eeq for odd $l$, and \beq \left(\frac
{dR_{nlj}}{dr}\right)_{r=R_c}=0, \label{bce} \eeq for even ones. Let
us denote it as BC1. The use of this version of the boundary
conditions has been partly justified by physical considerations in
\cite{NV}, but the dependence of the results on the particular
choice  has never been discussed in detail. It is the purpose of the
paper to study this problem at a quantitative level and to establish
the corresponding uncertainty, which is inherent to the WS method
applied to neutron star crust. For this aim, we compare results
obtained for the BC1 with those found for an alternative kind of the
boundary conditions (BC2) when Eq.~(\ref{bco}) is valid for even $l$
whereas Eq.~(\ref{bce}), for odd ones. In principle, two additional
kinds of the boundary conditions exist when Eq.~(\ref{bco}) or
Eq.~(\ref{bce}) is used for any $l$. As it was noted  in \cite{NV},
these versions have an obvious drawback for the case of the neutron
star inner crust since they lead to an unphysical irregular behavior
of the neutron density $\rho_n(r)$ in vicinity of the point $r=R_c$.
Indeed, $\rho_n(r)$ vanishes in this point in the first case and has
a maximum in the second one. On the contrary, $\rho_n(r)$ is almost
constant nearby the point $r=R_c$ in the case of the BC1 or BC2
kinds of the boundary conditions. It should be noted that the
pairing effects were not taken into account in \cite{NV} since  it
was supposed that they are not important for the structure of the
crust. The reason of such an assumption is the rather small
contribution of the pairing effects to the total binding energy of
the system under consideration. Recently, we have generalized the
approach by Negele and Vautherin to describe the inner crust by
explicitly including the neutron and proton pairing correlations
\cite{crust1,crust2,crust3,crust4} in a self-consistent way. It
turned out that in the whole interval of $\rho$ the equilibrium
configuration ($Z,R_c$) changes significantly due to pairing.
 To explain this effect, it is instructive
to analyze the $\beta$-stability condition (\ref{mu}). Since
electrons in the inner crust of a neutron star are
ultra-relativistic, the following relation is valid: $\mu_e \simeq
(9\pi Z/4)^{1/3}/R_c$. By substituting it into Eq.~(\ref{mu}), one
finds

\beq
Z \simeq \frac {4} {9 \pi} (\mu_n-\mu_p)^3 R_c^3.
\label{Zc}
\eeq

The influence of pairing on the chemical potentials $\mu_n$ and
$\mu_p$ is much stronger than that on the total binding energy.
Their variation may be of the order of the gap value $\Delta
\simeq$1--2$\;$MeV. Such a variation of $\mu_n$ or $\mu_p$ may lead
to a sizable change of the equilibrium value of $Z$ because the
difference ($\mu_n - \mu_p$) is raised to the third power in
Eq.~(\ref{Zc}). The estimate of the change of $Z$ induced by this
variation is as follows: $\delta Z{=}3Z\delta(\mu_n-\mu_p)/
(\mu_n-\mu_p)$. For average values of $k_{\Fs}$, the difference
$\mu_n-\mu_p \simeq 50 \div 70\;$MeV, hence $\delta Z$ could reach
few units of $Z$. An additional change of the $Z$ value may appear
due to a variation of $R_c$. Besides, as it is shown in
\cite{crust3,crust4},  the binding energy $E_B$ is rather flat
function of $Z$ and different local minima $E_B(Z)$ have often close
values of $E_B$. Therefore their relative position may change after
switching off the pairing since in general the corresponding
contribution to $E_B$ is an irregular function of $Z$. Such a
situation does often occur within the WS approach, especially for
high density values, due to the shell-type effects in the
single-particle neutron spectrum. An example is discussed in the
paper.

\section{Brief description of the method}

We used the generalized energy functional method  \cite{Fay} which
incorporates the pairing effects into the original Kohn-Sham
\cite{KS} method. In this approach, the interaction part of the
generalized energy functional depends, on equal footing, on the
normal densities $\rho_n , \rho_p$, and the abnormal ones, $\nu_n,
\nu_p$, as well: \beq E_{\rm int} = \int d {\bf r} {\cal E}_{\rm
int}(\rho(\bf r ),\nu({\bf r})), \label{GEF} \eeq where ${\cal
E}_{\rm int}$ is the energy functional density. It is the sum of
two components, the normal and the anomalous (superfluid) ones:
\beq {\cal E}_{\rm int} = {\cal E}_{\rm norm}(\rho_{\tau})+ {\cal
E}_{\rm an}(\rho_{\tau}, \nu_{\tau}), \label{Func} \eeq where
$\tau=n,p$ is the isotopic index. Just as in the Kohn-Sham method,
the prescription $m^*=m$ holds to be true. To describe the central
part of a WS cell with the nuclear cluster inside we used the
phenomenological nuclear energy functional ${\cal E}^{\rm ph}$ by
Fayans et al. \cite{Fay} which describes properties of the
terrestrial atomic nuclei with high accuracy. For describing
neutron matter surrounding the cluster we  used a microscopic
energy functional ${\cal E}^{\rm mi}$ for neutron matter based on
the Argonne NN potential  v$_{18}$ \cite{v18}. The ansatz of
\cite{crust3,crust4} for the complete energy functional is  a
smooth matching of the phenomenological and the microscopic
functionals at the cluster surface:

\beq {\cal E}(\rho_{\tau}({\bf r}),\nu_{\tau}({\bf r})) = {\cal
E}^{\rm ph}(\rho_{\tau}({\bf r}),\nu_{\tau}({\bf r})) F_m( r)+
{\cal E}^{\rm mi}(\rho_{\tau}({\bf r}),\nu_{\tau}({\bf r}))(1 -
F_m(r)), \label{tot} \eeq where the matching function $F_m(r)$ is
a two-parameter Fermi function:

\beq F_m(r)=(1+\exp((r-R_m)/d_m))^{-1}. \label{match}
\eeq
\noindent
Eq. (\ref{tot}) is applied  both to the normal and to the anomalous
components of the energy functional. After a detailed analysis, the
matching parameters were chosen as follows. The diffuseness
parameter was taken to be equal to $d_m{=}0.3\;$fm for any value of
the average baryon density of the inner crust and for any
configuration ($Z,R_c$). As to the matching radius $R_m$, it should
be chosen anew in any new case, in such a way that the equality \beq
\rho_p(R_m)=0.1 \rho_p(0) \label{match1} \eeq holds. In this case,
on one hand, for $r <\; R_m$ neutrons and protons coexist inside the
nuclear-type cluster, and the use of a realistic phenomenological
energy functional seems reasonable. On the other hand, at $r >\;
R_m$ one can neglect the exponentially decaying proton ``tails'' and
consider the system as a pure neutron matter for which an adequate
energy functional microscopically calculated can be used. The same
matching parameters  were used for normal and anomalous parts of
(\ref{tot}). As far as practically all the protons are located
inside the radius $R_m$, the matching procedure concerns, in fact,
only neutrons, protons being described with the pure
phenomenological nuclear functional. It is worth to mention that for
neutron matter region, the ansatz is, in fact, the LDA for the
microscopic part of the generalized energy functional. As it is
commonly known, the LDA works well only provided the density is
smoothly varying, whereas it fails in the surface region with a
sharp density gradient.  The above choice of the matching procedure
and the values of the parameters guarantees that this region of a
sharp density variation  is mainly governed by the phenomenological
nuclear part of the energy functional which ``knows how to deal with
it''. For the microscopic part of the normal component of the total
energy functional (\ref{tot}) we follow to refs.
\cite{crust3,crust4} and take the equation of states of neutron
matter calculated in \cite{B-V} with the Argonne v$_{18}$ potential
on the basis of Brueckner theory, taking into account a small
admixture of 3-body force. Its explicit form could be found in the
cited articles. The microscopic part of the anomalous component of
the generalized energy functional in \cite{crust3,crust4} was
calculated for the same v$_{18}$ potential within the
Bardeen-Cooper-Schrieffer (BCS) approximation.

\section{Comparison of two kinds of boundary conditions,
the BC1 versus BC2.}

 In the calculations of \cite{crust1,crust2,crust3,crust4}
the boundary condition by Negele and Vautherin, BC1, was used.
 Here we repeat the
analysis for the case of the boundary condition BC2. Results for the
binding energy per a nucleon, $E_B$, are shown in Fig.~1.

\begin{figure}
\centerline{\includegraphics [height=180mm,width=100mm]{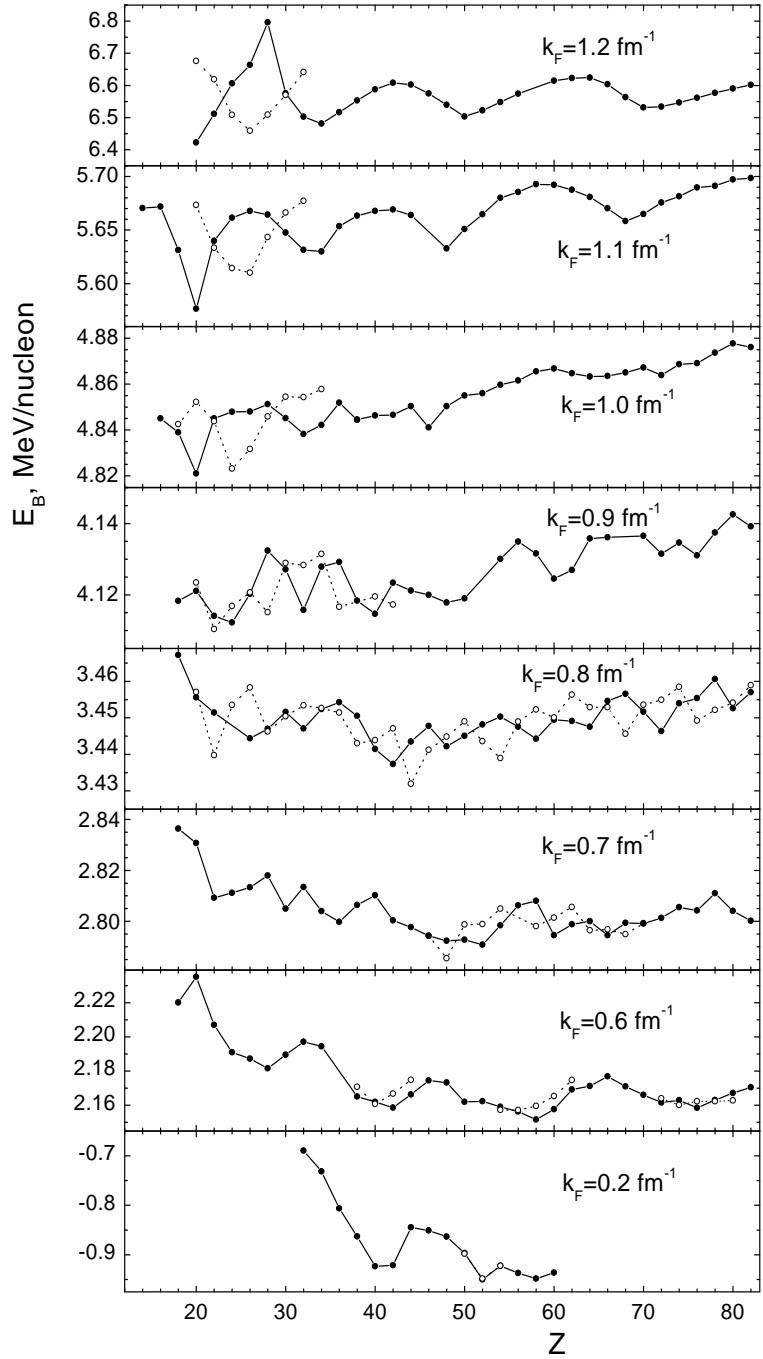}}%
\vspace{0mm} \caption{ Binding energy per a nucleon for various
$k_{\Fs}$ in the BC1 case  (solid circles connected with the solid
lines) and the BC2 one (open  circles connected with the dotted
lines).}
\end{figure}

It is worth to mention that calculations for larger values of
$k_{\Fs}$ should be considered as optional as far in this case the
spaguetti phase, evidently, is realized. This is true for
$k_{\Fs}{=}1.2\;$fm$^{-1}$ \cite{Oyam,Mag} and, maybe, for
$k_{\Fs}{=}1.1\;$fm$^{-1}$ \cite{Mag}. Just as in
\cite{crust3,crust4} only even values of Z are used. The detailed
comparison is made for $k_{\Fs}{=}0.8\;$fm$^{-1}$. Although the two
curves $E_B(Z)$ are quite different, the positions of local minima
for BC1 and BC2 are close to each other, the distance being equal to
2 or 4 units of Z. What is of primary importance, the relative
position of local minima for BC2 is the same as for BC1. In
particular, the positions of the absolute minimum almost coincide
($Z{=}52$ for BC1 and $Z{=}54$ for BC2). These observations permit
us to simplify calculations for other values of $k_{\Fs}$. In the
case of  BC2, we limit ourselves mainly with the analysis of a
vicinity of the absolute minimum for BC1. The neighborhood of other
local minima was analyzed only in the case if they have values of
$E_B(Z)$ close to that corresponding to the absolute minimum. It
turned out that there is no value of $k_{\Fs}$  for which the
relative position of a local minimum and of the absolute one for BC1
and BC2 is different. In addition to systematic calculations for
$k_{\Fs}{=}0.6 \div 1.2\;$fm$^{-1}$, we made an extra one for a
small density, $k_{\Fs}{=}0.2\;$fm$^{-1}$, in vicinity of the
neutron drip point. In the last case, two curves corresponding to
BC1 and BC2 practically coincide. For all other values of $k_{\Fs}$
the absolute minima are shifted by 2, 4 or even 6 units of $Z$.
Comparison of different properties of the equilibrium configuration
of the WS cell for various values of $k_{\Fs}$ in the case of BC1
and BC2 is presented in  Table 1.

\begin{table}[!b]
\caption{Comparison of properties of equilibrium configurations of
the WS cell for two different kinds of the boundary condition}

\bigskip
\begin{center}
\begin{tabular}{|c|c|c|c|c|c|c|c|c|c|c|}
\hline
  $k_{\rm F}$, & \raisebox{-6pt}{$Z$} &
  \multicolumn{2}{|c|}{$R_{\rm c},\;$fm}\rule{0pt}{14pt}&
  \multicolumn{2}{|c|}{$E_B,\;$MeV}&
  \multicolumn{2}{|c|}{$\mu_{\rm n},\;$MeV}&
  \multicolumn{2}{|c|}{$\Delta_{\rm F},\;$MeV}&
  {$\Delta_{\rm inf}$,}\\
\cline{3-10}
 \rule{0pt}{13pt}${\rm fm^{-1}}\!$  && BC1 & BC2 & BC1 & BC2 & BC1 & BC2 & BC1 &  BC2 &MeV\\
\hline
 0.2  & 52 & 57.18 & 57.10 &-0.9501 &-0.9483 & 0.1928 & 0.1942 & 0.04 & 0.05 & 0.40\\
\hline
 \raisebox{-6pt}{0.6} & 58 & 37.51 & 37.48 & 2.1516 & 2.1596 & 3.2074 & 3.2226 & 1.92 & 1.89
 & \raisebox{-6pt}{2.42}\\
      & 56 & 36.97 & 36.95 & 2.1563 & 2.1572 & 3.2173 & 3.2193 & 1.91 & 1.89 &    \\
\hline
  \raisebox{-6pt}{0.7} & 52 & 32.02 & 32.04 & 2.7908 & 2.7989 & 3.9876 & 4.0107 & 2.30 & 2.25
  & \raisebox{-6pt}{2.76}\\
      & 48 & 31.16 & 31.14 & 2.7924 & 2.7856 & 4.0069 & 3.9873 & 2.29 & 2.32 &    \\
\hline
  \raisebox{-6pt}{0.8} & 42 & 26.90 & 26.91 & 3.4373 & 3.4471 & 4.8454 & 4.8561 & 2.56 & 2.45
  & \raisebox{-6pt}{2.93}\\
      & 44 & 27.29 & 27.30 & 3.4435 & 3.4319 & 4.8553 & 4.8198 & 2.53 & 2.56 &    \\
\hline
  \raisebox{-6pt}{0.9} & 24 & 20.26 & 20.30 & 4.1123 & 4.1169 & 5.7340 & 5.7986 & 2.64 & 2.51
  & \raisebox{-6pt}{2.92}\\
     &  22 & 19.87 & 19.70 & 4.1141 & 4.1104 & 5.7861 & 5.7170 & 2.62 & 2.54 &    \\
\hline
  \raisebox{-6pt}{1.0} & 20 & 16.69 & 16.90 & 4.8210 & 4.8522 & 6.8525 & 6.7424 & 2.02 & 2.52
  & \raisebox{-6pt}{2.68}\\
      & 24 & 18.29 & 18.22 & 4.8479 & 4.8231 & 6.8446 & 6.8920 & 2.52 & 2.29 &    \\
\hline
  \raisebox{-6pt}{1.1} & 20 & 14.99 & 15.33 & 5.5765 & 5.6733 & 7.4288 & 8.0446 & 1.32 & 2.32
  & \raisebox{-6pt}{2.26}\\
      & 26 & 16.75 & 17.08 & 5.6677 & 5.6100 & 7.9680 & 8.5398 & 2.28 & 2.02 &    \\
\hline
  \raisebox{-6pt}{1.2} & 20 & 13.68 & 13.95 & 6.4225 & 6.6762 & 8.5814 & 9.1898 & 1.21 & 1.56
  & \raisebox{-6pt}{1.66}\\
      & 26 & 15.21 & 14.89 & 6.6639 & 6.4587 & 9.0825 & 9.3413 & 1.25 & 0.86 &  \\[1mm]
\hline
\end{tabular}
\end{center}
\end{table}

 There are
two lines for every value of $k_{\Fs}$. The first one is given for
the $Z$ value corresponding to the minimum of $E_B$ in the BC1
case, the second one, for BC2. The only exception is
$k_{\Fs}{=}0.2\;$fm$^{-1}$ when these two values of $Z$ coincide.
In the last couple of columns, the average value $\Delta_{\Fs}$ of
the diagonal matrix element of the neutron gap at the Fermi
surface is given. The averaging procedure involves 10 levels above
$\mu_n$ and 10 levels below. For a comparison, the infinite matter
value $\Delta_{\rm inf}$ is given in the Table. It is calculated
within the BCS approximation for the density of neutron matter
corresponding to $k_{\Fs}$ value under consideration. So drastic
difference between $\Delta_{\rm inf}$ and $\Delta_{\Fs}$ values in
the case of $k_{\Fs}{=}0.2\;$fm$^{-1}$ is caused by the trivial
reason that this  $k_{\Fs}$ is in the vicinity of the neutron drip
point. Therefore the asymptotic neutron density which determines
mainly the $\Delta_{\Fs}$ value is significantly less than that
for the uniform matter distribution.
 One can see that the influence of the
boundary conditions  is enhanced at increasing values of $k_{\Fs}$.
Especially strong variation of  $\Delta_{\Fs}$ and $\mu_n$ values
takes place in the cases of $k_{\Fs}{=}1.1\;$fm$^{-1}$ and
$k_{\Fs}{=}1.2\;$fm$^{-1}$, see Fig. 2.

\begin{figure}[!b]
\centerline{\includegraphics [height=100mm,width=120mm]{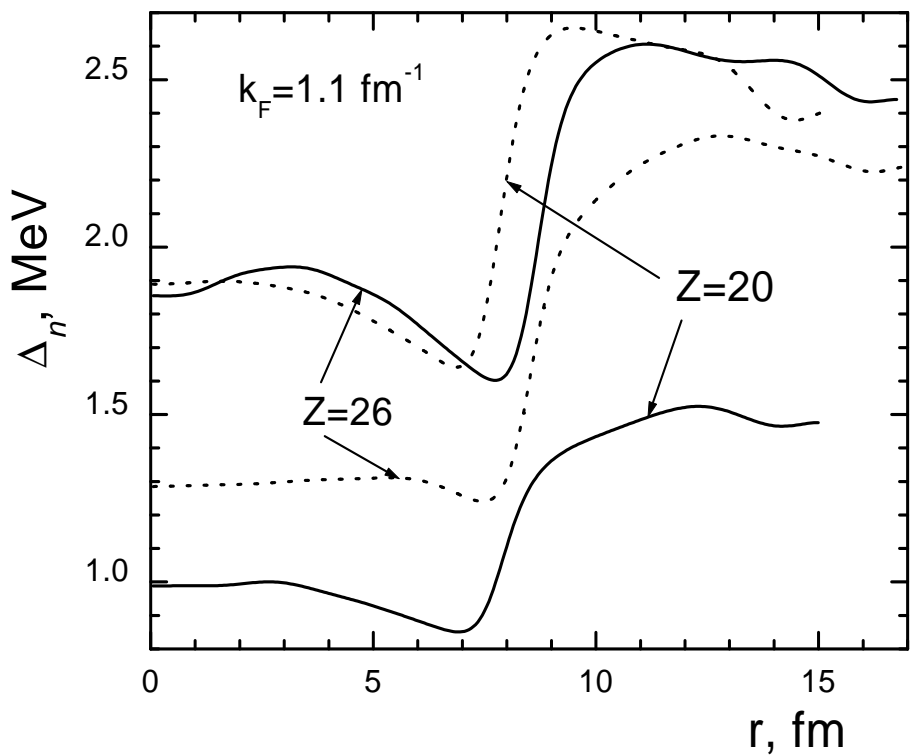}}%
\vspace{0mm} \caption{ The neutron gap for
$k_{\Fs}{=}1.1\;$fm$^{-1}$, $Z{=}20$ and $Z{=}26$, in the BC1 case
 (solid lines) and the BC2 one (dashed lines).}
\end{figure}

To illustrate the influence of the boundary conditions on the
neutron gap in the first case, the gap function $\Delta_n(r)$ is
drawn for both values of $Z$ and both kinds of the boundary
conditions. The strongest variation of the gap occurs in the case of
$Z{=}20$. To understand  the reason of such strong effect, we draw
the neutron single particle spectrum $\eps_{\lambda}$ for this value
of $Z$ in Fig. 3 for the BC1 case (the left half of the figure) and
the BC2 one (the right one). The position of the chemical potential
$\mu_n$ is shown with dots.

\begin{figure}[!t]
\centerline{\includegraphics [height=130mm,width=100mm]{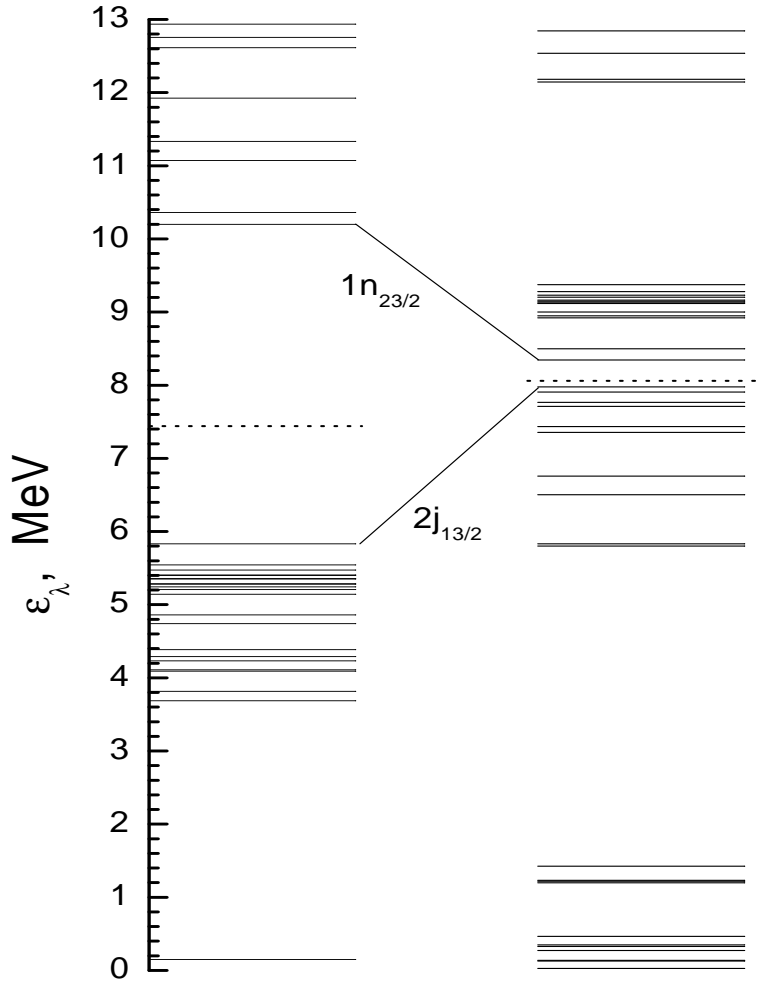}}%
\vspace{0mm} \caption{ The neutron single-particle spectrum
$\eps_{\lambda}$ for $k_{\Fs}{=}1.1\;$fm$^{-1}$, $Z{=}20$, in the
 BC1 case  (left) and the BC2 one (right).}
\end{figure}

The two spectra are absolutely different. The reason is the shift
$\Delta \eps_{\lambda}$  of each $\lambda$-level going from BC1 to
BC2. The value of this shift is approximately equal to a half of the
distance between two neighboring levels with the same ($l,j$),
 the sign of the shift being opposite for even and odd $l$. The
absolute value of the shift is proportional to $1/R_c^2$ and grows
at increasing values of $k_{\Fs}$. The corresponding shifts are
shown in Fig. 3 for two states, 2$j_{13/2}$ and 1$n_{23/2}$, which
are the neighbors of $\mu_n$ in the BC1 case. On average, the
spectrum is quite dense, however in both cases there is a shell
type structure with rather wide intervals between some neighboring
levels. If one deals with a big inter-level space in vicinity of
$\mu_n$, as in the BC1 case in Fig. 3, one usually obtains a dense
set of levels in this region when going to the opposite kind of
the boundary conditions. In the  BC2  case, big intervals are far
from the Fermi surface and do not influence significantly the
value of the neutron gap. On the contrary, in the BC1 case
$\mu_n$ is situated just inside such an interval that suppresses
the gap significantly. In principle, the neutron gap could vanish
if the interval was wider.

\begin{figure}[!h]
\vskip 0.3 cm
\centerline{\includegraphics [height=100mm,width=120mm]{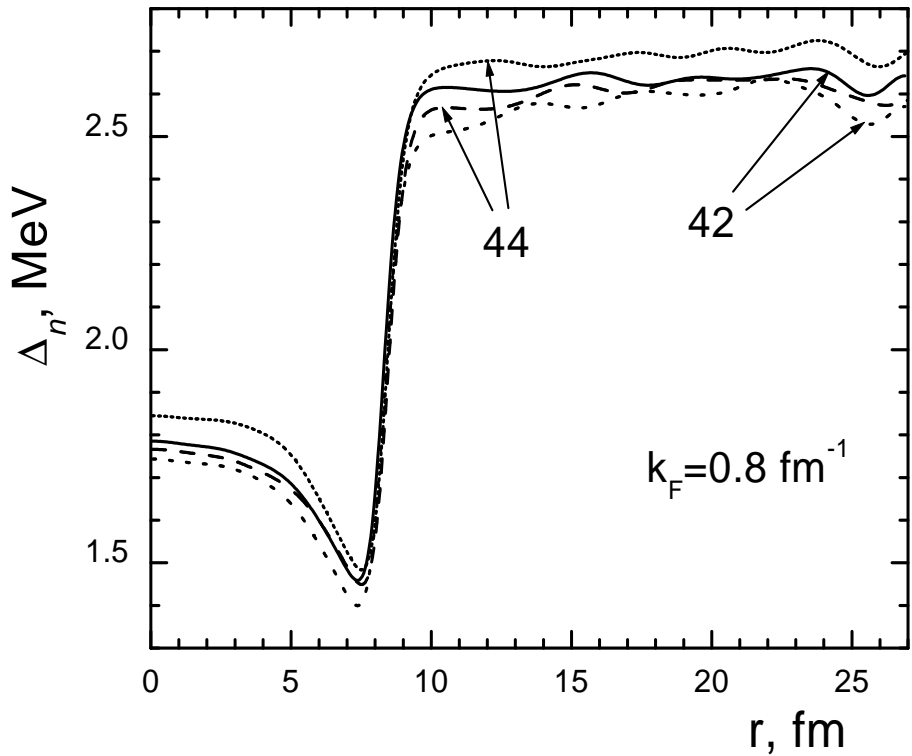}}%
\vspace{0mm} \caption{The neutron gap for
$k_{\Fs}{=}0.8\;$fm$^{-1}$, $Z{=}42$ and $Z{=}44$, in the BC1 case
 (solid lines) and the BC2 one (dashed lines).}
\end{figure}

As to the gap function itself, for intermediate densities, $k_{\Fs}
< 1\;$fm$^{-1}$, variations caused by the choice of the boundary
conditions are not significant. An example for the case of
$k_{\Fs}{=}0.8\;$fm$^{-1}$ is given in Fig. 4 where, just as in Fig.
2, four curves for $\Delta(r)$ are shown. The difference between any
couple of these curves is less, of course, than the accuracy of the
approach. Evidently, the most important uncertainty in the neutron
gap value comes from using the BCS approximation  without many-body
effects like screening, which, for infinite neutron matter,
overestimates $\Delta_n$ significantly \cite{crust2}. \par To be
definite, let us consider the ``self-consistent'' ($Z{=}42$ for
$k_{\Fs}{=}0.8\;$fm$^{-1}$~) gap function for the BC1 version of the
boundary conditions as the prediction of the WS method for
$\Delta(r)$ in the case of small and intermediate densities,
$k_{\Fs}< 1.0\;$fm$^{-1}$. Such a choice corresponds to that used in
\cite{crust4}. The ``normal'' situation with the gap in the case of
intermediate densities corresponds to much more regular
single-particle neutron spectra, than that in Fig. 3. For the case
of $k_{\Fs}{=}0.8\;$fm$^{-1}$ it is illustrated in Fig. 5. One can
see that, although here also there are some irregularities in the
inter-level distances, they are much smaller than the gap value
$\Delta \simeq 2.5\;$MeV. Therefore they don't influence the gap
equation significantly.

\begin{figure}
\centerline{\includegraphics [height=130mm,width=100mm]{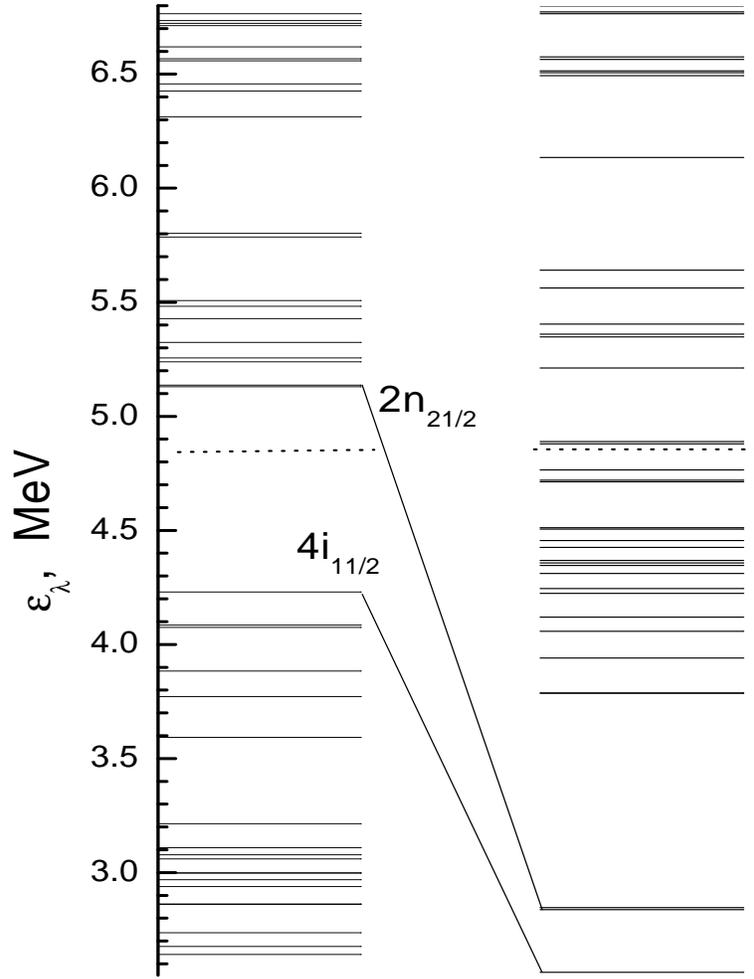}}%
\vspace{0mm} \caption{ The neutron single-particle spectrum
$\eps_{\lambda}$ for $k_{\Fs}{=}0.8\;$fm$^{-1}$, $Z{=}42$, in the
BC1 case (left) and the BC2 one (right).}
\end{figure}

In the case of bigger densities, as it is seen in Fig.~2, there is a
significant uncertainty in predictions for $\Delta(r)$ within the WS
method. As it was discussed above, it originates from the shell-type
structure of the neutron single-particle spectrum which appears in
the case of high $k_{\Fs}$ (small $R_c$) values. We consider this
effect as an artifact of the WS approximation which must disappear
(or almost disappear) in a more advanced approach with the
consistent consideration of the band structure. No doubts, if a
forbidden space between two bands in vicinity of $\mu_n$ in the band
structure calculation will appear, it should be much less than the
gap $\Delta$ value. Therefore the solution of the gap equation in
the realistic case should be closer to that in the WS approximation
for such a situation when there is no big inter-level interval close
to $\mu_n$. Returning to Fig. 2, these are the dotted line for
$Z{=}20$ and the solid one, for $Z{=}26$. Again the difference
between these two curves is not essential, and any of them could be
used as the prediction for $\Delta(r)$ in the case of
$k_{\Fs}{=}1.1\;$fm$^{-1}$. Strictly speaking, such a recipe is not
a self-consistent one within the WS method, but it looks reasonable
from the physical point of view.

\section{Discussion and conclusions.}

As we have seen,  there are internal uncertainties inherent to the
WS method applied to the neutron star inner crust which originate
from the kind of the boundary condition used.
  In the case of very small density nearby the neutron drip
 point the predictions of the BC1 and BC2 versions are practically
 identical.  At increasing density, with $k_{\Fs} \ge 0.6\;$fm$^{-1}$,
 the  uncertainty in the equilibrium value of $Z$ is between 2 and
 6 units, with the largest values at the largest $k_{\Fs}$. The
 uncertainty in the value of $R_c$ is, as a rule, about 1 fm and
 only for $k_{\Fs}{=}1.1\;$fm$^{-1}$ it turns out to be about 2
 fm. However, the value of these uncertainties is  smaller than the
variations of the equilibrium configuration ($Z,R_c$) connected with
 the pairing effects \cite{crust4}. For the case of small and
intermediate densities , $k_{\Fs} < 1\;$fm$^{-1}$, the uncertainty
in predictions for the gap function $\Delta(r)$ caused by a
particular choice of the boundary condition is also rather small.
 In the case of high densities, $k_{\Fs} \gsim 1\;$fm$^{-1}$, the
 most important uncertainty occurs in the value of the neutron gap
 $\Delta_n$. In the WS approximation, it originates from the shell
effect in the
 neutron single-particle spectrum which is rather pronounced in
 the case of big $k_{\Fs}$ and, correspondingly, smaller  $R_c$ values.
We consider this effect as an artifact of the WS method which should
disappear in a more consistent band structure calculation. We
suggest an approximate recipe to avoid this uncertainty for the gap
function $\Delta(r)$. We think that  the most important uncertainty
in the neutron gap value comes from the BCS approximation used. As
it is well known, in neutron matter the many-body corrections to the
BCS approximation reduce the value of $\Delta_n$ significantly
\cite{crust2}. In the next paper, we plan to take into account this
reduction effect for more realistic description of neutron
superfluidity in the inner crust of neutron stars.

\vskip 0.5 cm The authors thank  N.E. Zein for valuable discussions.
This research was partially supported by the Grant NSh-8756.2006.2
of the Russian Ministry for Science and Education and by the RFBR
grant 06-02-17171-a.

{}
\end{document}